\documentclass[a4paper,12pt]{article}
\usepackage{amsmath}
\usepackage{ulem}
\usepackage{calc}
\usepackage{vmargin}
\usepackage{amstext}
\usepackage{graphicx}
\usepackage{amsfonts}
\usepackage{amssymb}

\begin{document}

	\title{
		{\bf Contact germs and partial differential equations } }
	
	\author{ \bf O.V. Kaptsov
		\\ Institute of Computaional Modelling SD RAS,
		\\ Krasnoyarsk, Russia
		\\ E-mail: kaptsov@icm.krasn.ru}

	\date{}
	\maketitle
	
The article introduces contact germs that transform solutions of some partial differential equations into solutions of other equations.
Parametric symmetries of differential equations generalizing point and contact symmetries are defined. 
New transformations and symmetries may depend on derivatives of arbitrary but finite order. The stationary Schrödinger equations, acoustics and gas dynamics equations are considered as examples. 


\section{Introduction}

The theory of point symmetries is now widely used to find solutions to differential equations. There are many monographs describing the theory and its applications in detail \cite{Ovs,Ibragimov,Olver, Bluman}.
In the second half of the last century, generalizations of point symmetries began to appear. Among them we mention, first of all, higher, conditional and asymptotic symmetries \cite{Vin,And,Win,Gaeta}.
These symmetries were found for the KdV, sine-Gordon, KP and others equations.
Higher symmetries make it possible to identify equations that have a large number of solutions expressed by means of elementary and special functions.
However, it is rarely possible to directly apply higher symmetries to construct solutions of nonlinear equations. On the other hand, operator symmetries of the first and second orders are actively used to find solutions of linear equations \cite{Miller,Kalnins}. In particular, symmetries allow us to find coordinate systems in which the equations admit the separation of variables. 

Most problems in point symmetry theory are well understood, but there are open questions.
These include, for example, the problem of constructively describing non-isomorphic subalgebras of a Lie algebra admitted by  differential equations.
Note that the transformation groups, in the general case, are local. Obviously, there is no need to distinguish between the neighborhoods of the point in which these groups are defined. Therefore, it makes sense to immediately consider the germs of transformations, which is what we do in this article.
 
 The article has the following structure. 
 In the second section, basic concepts are introduced such as prolonged spaces, contact modules of differential 1-forms, differential families of germs, the differential ring of infinite germs and ideals generated by differential families.
 Differential equations are defined using differential families of germs.
In the next section, we introduce contact germs and  prove the lifting lemma, which allows one to construct prolongations of germs to contact ones. The found prolongation formula generalizes the classical one for point mappings. 
  The fourth section gives examples of contact germs that transform solutions of some differential equations into solutions of other equations. As examples, we consider the one-dimensional acoustic equation and the three-dimensional Schrödinger equation.
  In the next section, parametric germs and symmetries are introduced. The new symmetries generalize classical point symmetries. Transformations defining symmetries can depend on derivatives above the first order and, as a rule, are irreversible.
  This allows us to avoid the restrictions of B\"{a}cklund's theorem \cite{Storm}. Examples of such symmetries for one parabolic equation and for a system of gas dynamics equations are given.

\section{Basic concepts}

Let $\mathbb{N}$ be the set of non-negative integers, $\mathbb{R}$ be the field of real numbers, and 
$P_{i,n}$ is the space of homogeneous polynomials of degree $i$ in $n$ variables with coefficients in $\mathbb{R}$.
We call the set $J^k = \mathbb{R}^m\times \prod_{i=1}^{k} P_{i,n} $ the space of $k$-jets,
and $P^k = \mathbb{R}^n\times J^k$  the $k$ times prolonged space $P^0 = \mathbb{R}^{n+m}$ or
the $k$-th prolongation  of $P^0$. 
Let  $P^{\infty}$ be the space  $P^0\times \prod_{i=1}^{\infty} P_{i,n} ,$
$\pi^j_i$ the projection of $P^j$ onto $P^i$, and $\pi_i$ the projection of $P^{\infty}$ onto $P^i$.
The coordinate functions on $P^{\infty}$ are denoted by $x_k, u^l_{\alpha}$, where
$k=1,\dots,n$, $l=1,\dots, m$, $\alpha\in\mathbb{N}^n$.

Let $a_i=\pi_i(a)$ be the projection of a point $a\in P^{\infty}$, $\mathcal{F}_{a_i}$  the ring of germs of smooth functions at point $a_i$  for every $i\in \mathbb{N}$, and
$\nu^j_i: \mathcal{F}_{a_i}\longrightarrow \mathcal{F}_{a_j} $  the the canonical embedding  ($i\leq j$).
The family of rings $(\mathcal{F}_{a_i})_{i\in \mathbb{N}} $ together with the embeddings $\nu^j_i$ forms a directed system \cite{Rotman}.
The direct limit of $\mathcal{F}_{a}^\infty =\underrightarrow{\lim} \mathcal{F}_{a_i}$
is called the ring of $\infty$-germs at point $a$, with embeddings 
$\nu_i : \mathcal{F}_{a_i}\rightarrow \mathcal{F}_{a}^\infty$.
We introduce the total derivatives (derivation operators) 
$$ D_{x_k}(f) = \frac{\partial f}{\partial x_k} + 
\sum_{\substack{1\leq j\leq m \\\ \alpha\in \mathbb{N}^n }} \frac{\partial f}{\partial u^j_\alpha}
u^j_{\alpha +1_k} \quad \forall f\in\mathcal{F}_a, \quad k=1,\dots n,$$
where $1_k$ is an $n$-dimensional vector with all coordinates zero except the $k$-th coordinate, which is equal to $1$. Then $\mathcal{F}_a$ becomes a differential ring.

We denote by $\Omega_{a_i}$ the module over the ring $\mathcal{F}_{a_{i+1}}$ generated by the differentials $dx_i, du^j_\alpha$, where $i=1,\dots,n$, $j=1,\dots, m$, $\alpha\in\mathbb{N}^n$.
The elements of this module are called germs of differential 1-forms.
For any $r,s\in\mathbb{N}$ such that $r\leq s$, there are linear maps $\eta^s_r : \Omega_{a_r}\longrightarrow \Omega_{a_s}$,
given by the formulas $\eta^s_r(fdg)=\nu^s_r(f)dg$.  This is how a directed system $\{\Omega_{a_r},\nu^s_r \}$ arises, which has a direct limit $\Omega_a^\infty = \underrightarrow{\lim}\Omega_{a_r}$.

 \noindent
 {\bf Definition.} A submodule $C_{a_k}$ of the module $\Omega_{a_k}$ generated by differential forms  of the form 
 $$\omega^j_\alpha = du^j_\alpha - \sum_{1\leq i\leq n} u^j_{\alpha+1_i}dx_i, 
 \qquad |\alpha|\leq k $$
 is called contact.

 \noindent
{\bf Definition.} The finite set $E=\{f_1,\dots,f_l\}\subset\mathcal{F}_{a_k}$ is called a differential family, and the expression 
\begin{equation} \label{f1=0} 
f_1=\cdots=f_l=0  
\end{equation}
by a system of differential equations. 
 
Any differential family $E$ generates a differential ideal $<E>$ of the ring $\mathcal{F}_{a}^\infty$. The elements of an ideal are represented as a sum of germs of the form $g_\alpha D^\alpha f_i$, where $\alpha=(\alpha_1,\dots,\alpha_n)$, $D^\alpha = D_{x_1}^{\alpha_1}\cdots D_{x_n}^{\alpha_n} $ , $f_i\in E$, $g_\alpha\in \mathcal{F}_{a}^\infty$.
The germ of a smooth mapping $f:\mathbb{R}^{k_1} \longrightarrow  \mathbb{R}^{k_2}$ at a point $p\in\mathbb{R}^{k_1}$ is denoted by $f: (\mathbb{R}^{k_1},p)\longrightarrow \mathbb{R}^{k_2}$.

 \noindent
{\bf Definition.} If $s_0 :(\mathbb{R}^{n},c)\longrightarrow \mathbb{R}^{m}$ is a germ, then its $k$-th prolongation is the germ 
$s_{k} :(\mathbb{R}^{n},c)\longrightarrow P^k$ given by  $s_{k}=(q_0,q_1,.. .,q_k)$, where $q_0=s_0$,
$q_i=\frac{\partial^\alpha(s_0)}{\partial x_1^{\alpha_1}\dots\partial x_n^{\alpha_n} }$, $|\alpha|=i>0$. An infinite prolongation of the germ $s_0$ is the sequence
$s_{\infty}=(q_i)^{\infty}_{i=0}$

 \noindent
 {\bf Definition.} Let $E=(f_i)_{i=1}^l\subset\mathcal{F}_{a_k}$ be a differential family. A germ
 $s_0 :(\mathbb{R}^{n},c)\longrightarrow \mathbb{R}^{m}$ is called a solution to the system (\ref{f1=0}),
 if its $k$-th prolongation annihilates the family, i.e. $f_i\circ s_k =0$ for $i=1,\dots,n$.
 We will say that $s_{\infty}$ annihilates the ideal $<E>$ if $f\circ s_{\infty}= 0$\
 $\forall f\in <E>$.
 
 {\it Remark.} The germ $s_k$ annihilates the family $E$ if and only if $s_{\infty}$ annihilates the ideal $<E>$. This follows from the well-known property of solutions of the differential equations \cite{Ovs}:
 if $s_k$ annihilates $f$, then $s_{k+1}$ annihilates $D_{x_i}(f)$ for any $i=1,...,n$.

\section{Contact germs}

In this section we introduce contact germs of smooth maps of prolonged spaces acting on solutions of differential equations.

 Let $\phi : (P^j,a_j)\longrightarrow P^i$ be a germ, where $j\geq i$ and $\phi(a_j)=b_i$.
It induces a ring homomorphism 
$ \phi^{\star} :  \mathcal{F}_{b_i}  \longrightarrow \mathcal{F}_{a_j} $ with 
$$\phi^{\star} (f)= f(\phi) \quad \forall f\in \mathcal{F}_{b_i}$$
and a linear map of the modules 
$\psi^{\star} : \Omega_{b_i}\longrightarrow \Omega_{a_j}$, given by
$$  \psi^{\star}(fdg) =  \phi^{\star} (f)d\phi^{\star}(g)   
\quad \forall f,g\in \mathcal{F}_{b_i}$$

Matrix $\mathfrak{D}\phi$ with elements
$D_{x_i}(\phi_k)$, 
where $\phi_k$ is the $k$-th component of the germ and $i=1,\dots n$,
is called the total Jacobian matrix of the germ $\phi$.

\noindent
{\bf Definition.} A germ $\phi : (P^j,a_j)\longrightarrow (P^i,b_i)$ is called contact of order $i$ if the mapping $\psi^{\star}$ preserves contact modules, i.e. $\psi^{\star}(C_{b_i})\subset C_{a_j}$.

The following lifting lemma describes a way of constructing contact germs. The classical statement can be found in $\cite{Ovs,KaptsovBook}$.

\noindent
{\bf Lemma 1.} Let $\Phi_{0} : (P^k,a_k)\longrightarrow P^0$ be a germ of the form
$$ y_i = f_i(x_1,\dots,x_n,u^1,\dots,u^m_{\alpha}) ,\qquad 1\leq i\leq n,
\ |\alpha|\leq k , $$
$$ \tilde{u}_j = g_j(x_1,\dots,x_n,u^1,\dots,u^m_{\alpha}) ,
\qquad 1\leq j\leq m, $$
and the total Jacobian matrix $\mathfrak{D}f $ is invertible.
Then for any $s\geq 1$ there is a contact germ
$\Phi_{s} : (P^{k+s},a_{k+s}) \longrightarrow P^s$ of the form $(v_0,\dots,v_s)$, where $v_0=\Phi_{0} $,
$v_1= (\mathfrak{D}g)(\mathfrak{D}f)^{-1}$, and the remaining components are given by the formula
\begin{equation} \label{vi} 
	v_{i+1}=(\mathfrak{D}v_i)(\mathfrak{D}f)^{-1}       .       
\end{equation}

\noindent
{\it Proof.} 
Let us introduce the notations $f=(f_1,\dots,f_n)$, $g=(g_1,\dots,g_m)$, $dx=(dx_1,\dots,dx_n)$, $du_{\alpha}=(du ^1_\alpha,\dots,du^m_\alpha)$,
$u_{\alpha+1}=\mathfrak{D}u_\alpha$.
To construct a first-order contact germ, we need to find matrices $v_1, B_0,\dots,B_k$ that satisfy the condition
\begin{equation} \label{dg} 
	dg - v_1df = B_0(du-u_1dx) + \cdots + B_k(du_\alpha -u_{\alpha+1}dx) ,\qquad |\alpha|=k .
\end{equation}
The left side of the last formula is written as
$$ g_xdx +g_udu+\dots+g_\alpha du_\alpha -
v_1(f_xdx +f_udu+\dots+f_\alpha du_\alpha ) , $$
where $g_x=(\frac{\partial g_j}{\partial x_i}), \ (\frac{\partial g_j}{\partial u^i}), \dots, f_{u_{\alpha}}=( \frac{\partial f_j}{\partial u^i_\alpha})$ are standard Jacobian matrices.

From the coefficients of $dx, du,...,du_{\alpha}$ in (\ref{dg}) we get
$$ g_x - v_1f_x +B_0u_1+ \dots+B_ku_{\alpha+1} =0 ,  \qquad |\alpha|=k $$
$$  g_u -v_1f_u=B_0,\quad \cdots , \quad g_{u_{\alpha}} -v_1f_{u_{\alpha}}=B_k .  $$
Substituting $B_0,...,B_k$ into the first equation of this system, we have
$$ \mathfrak{D}g = v_1\mathfrak{D}f . $$
By condition, the matrix $\mathfrak{D}f$ is invertible.
This means that the prolongation formula has the form
\begin{equation} \label{v_1}
	v_1 = \mathfrak{D}g (\mathfrak{D}f)^{-1}  .
\end{equation}
Thus, a contact germ $\Phi_1=(v_0,v_1)$ of the first order is obtained. The formula (\ref{vi}) is proved in a similar way.

The germ $\Phi_k$ constructed in the previous lemma will be called the $k$-th prolongation of the germ $\Phi_0$. We will call an infinite sequence $\Phi_\infty=\{v_i\}_{i\geq 0}$ a $\infty$-contact germ.

Let us define the homomorphism $\Phi_\infty^\star : \mathcal{F}_b^\infty \longrightarrow \mathcal{F}_a^\infty $ by the formula
 $\Phi_\infty^\star(f) = f(\Phi_\infty)$ 
 for any $f\in \mathcal{F}_b^{\infty}$.

\noindent
{\bf Proposition.} Let $E_1\subset \mathcal{F}_{a_k}$ and $E_2\subset \mathcal{F}_{b_k}$ be differential families, 
$\Phi_\infty : P_a^\infty \longrightarrow P^{\infty}_b$ 
be an $\infty$-contact germ, and $s_\infty$ annihilates the ideal $<E_1>$.
If $\Phi_\infty^\star(<E_2>)\subset <E_1>$, then $\Phi_\infty\circ s_\infty$  annihilates the ideal $<E_2>$.

Indeed, according to the conditions, $g\circ \Phi_\infty\in<E_1>$ for any $g\in<E_2>$ and $f\circ s_\infty= 0$ $\forall f\in<E_2>$ . Then the following is true
$$ g\circ \Phi_\infty\circ s_\infty =0 .$$
Thus, contact germs act on solutions of differential systems.

\noindent
{\bf Definition.} Differential families $E_1\subset \mathcal{F}_{a_k}$ and $E_2\subset \mathcal{F}_{b_k}$ are called equivalent if there exist $\infty$-contact germs
$\Phi_\infty : P_a^\infty \longrightarrow P^\infty_b$ ,
$\Psi_\infty : P_b^\infty \longrightarrow P^\infty_a$ such that $\Phi_\infty^\star(<E_2>)\subset<E_1>$ and $\Psi_\infty^\star(< E_1>)\subset<E_2>$.

\noindent
{\bf Definition.} Let $E=(f_i)_{i=1}^l\subset\mathcal{F}_{a_k}$ be a differential family. If there are $\infty$-contact germs
$\Phi_\infty : P_a^\infty \longrightarrow P^\infty_a$
such that $\Phi_\infty^\star(<E>)\subset<E>$, then $\Phi_\infty$ is called the symmetry of the system
$$ f_1 = 0 , \cdots, f_l = 0 . $$

\section{Actions of germs on solutions }

This section provides examples of contact germs acting on solutions of differential equations.
For convenience, classical notations are used.
So instead of $u^j_\alpha $ we will use the notation
$\frac{\partial^\alpha u^j}{\partial x_1^{\alpha_1}\cdots\partial x_n^{\alpha_n}}$ or even shorter ones like $u_{tt}, u_{xx}$ .

As a first example, consider the equation
\begin{equation} \label{u_tt}
	u_{tt} = (x^2 u_x)_x ,
\end{equation}
such equations were studied in \cite{Bre} in connection with applications to acoustics. 
  
  We will seek germs of the form
  \begin{equation} \label{tau}
  	\tau = t, \quad y=h(x), \quad v=f(x)u_x +g(x)u
  \end{equation}
  whose prolongations  are symmetries of the equation (\ref{u_tt}).
 Here $f, g, h$ are germs depending on $x$.
According to Lemma 1, the first and second prolongations of the germ (\ref{tau}) are given by the formulas 
$$  v_t =  fu_{tx}+gu_t , \qquad 	v_y = 
\frac{D_x(fu_x+gu)}{ h^{\prime}}  ,$$
$$v_{tt} = fu_{ttx}+gu_{tt}\ ,\qquad v_{yy} = \frac{D_x(v_y)}{h^{\prime}} \ . $$
Substituting these formulas into the left side of the equation
$$	v_{tt}-(y^2v_y)_y =0 , $$
we get an expression containing $u_{tt}, u_{ttx}$. We calculate these derivatives using the equation (\ref{u_tt}). As a result, we have a new expression, which is a polynomial of the first degree in $u_{xxx}, u_{xx}, u_x, u$.
Equating the coefficients of the polynomial to zero, we obtain a system of ordinary differential equations for the functions $f, g, h$.
  Solving the obtained system, we find the transformation
 of the form 
 $$ \tau = t, \quad y=\frac{c_0}{x}, \quad v=c_1 x^2u_x +c_2 x u , $$
 where $c_0, c_1, c_2$ are arbitrary constants. 

\noindent
{\it Remark.}  Equation (\ref{u_tt}) by substitution 
$$ u=\frac{1}{\sqrt{x}} w(t,\ln x)$$       
is reduced to the Klein-Gordon equation
$$ w_{tt} = w_{zz} -\frac{1}{4}w . $$ 

Let us now proceed to construct the transformations connecting
the three-dimensional stationary Schrödinger equation 
 \begin{equation} \label{Shre}
	\Delta \psi + U(x_1,x_2,x_3)\psi =0 ,
\end{equation}
 with Laplace's equation.

We will need the following statement \cite{KaptsovBook}.
Given a linear partial differential equation 
\begin{equation} \label{U2x}
	u_{xx} +H(x)u + Lu = 0,
\end{equation}
where $L$ is an operator of the form
$$ L = \sum_{|\alpha|\geq 0}^{k} a_\alpha(y)   
\frac{\partial^\alpha }{\partial y_1^{\alpha_1}\cdots\partial y_n^{\alpha_n}} , \qquad 
\alpha= (\alpha_1,\dots,\alpha_n) ,          $$
then the Euler-Darboux transformation
\begin{equation} \label{E-D}
	v = u_{x} - \frac{h^\prime (x) }{h(x)}u
\end{equation}
maps solutions of the equation (\ref{U2x}) into solutions of the equation
$$ v_{xx} +H_1(x)v + Lv = 0, $$
if $h$ satisfies the ordinary differential equation
\begin{equation} \label{h}
	h^{\prime\prime} +(H(x)+c)h = 0, \qquad c\in\mathbb{R} ,
\end{equation}
and $H_1$ has the form
$$ H_1 = H + 2 (\ln h)^{\prime\prime} . $$

Consider Laplace's equation
$$ u_{xx} + u_{yy} +u_{zz} = 0 $$
and apply the Euler-Darboux transformation (\ref{E-D}) to it.
Then the function $v(x,y,z)$ satisfies the Schrödinger equation of the form 
\begin{equation} \label{vxx}
	v_{xx} +v_{yy} +v_{zz} + 2 (\ln h)^{\prime\prime}v = 0,
\end{equation}
where $h$ is a solution of the equation
$$ h^{\prime\prime} +c h =0, \quad c\in\mathbb{R} .$$
The solutions of the last equation have the form
$$ h= c_1 x + c_2 , \quad  (c=0 ) , $$
$$ h= c_1 \exp(kx) + c_2 \exp(-kx), \quad  (c=-k^2<0) , $$
$$ h= c_1 \sin(kx) + c_2 \cos(-kx), \quad  (c=k^2>0) . $$
As a result, the function $H_1$ has one of three forms
$$  \frac{-2 c_1^2}{(c_1x+c_2)^2} ,\qquad \frac{ 8c_1c_2k^2}{(c_1 \exp(kx) + c_2\exp(-kx))^2} ,   \qquad \frac{- 8c_1c_2k^2}{ (c_1 \sin(kx) + c_2 \cos(-kx))^2 }    . $$
Next, we apply the Euler-Darboux transformation to the equation (\ref{vxx})
$$ v^1 = v_{y} - v\frac{h^\prime (y)}{h(y)} $$
and we arrive at the Schrödinger equation
$$ \Delta v^1 + (H_1(x)+H_2(y))v^1 =0 , $$
where $H_2(y)$ has one of the three forms indicated above, with $x$ replaced by $y$.
If we now perform the Euler-Darboux transformation on $z$, we obtain the equation
$$ \Delta v^2 + (H_1(x)+H_2(y)+H_3(z))v^2 =0 . $$
To the last equation we can again apply the Euler-Darboux transformations in $x$, $y$, $z$ and obtain new equations with new potentials. In this case, it is necessary to find solutions to the corresponding equations of the form (\ref{h}).

 Another way to transform the Schrödinger equation (\ref{Shre}) is to apply a group of three-dimensional conformal transformations to it, leaving the Laplace equation invariant.
For example, the  inversion transformation
$$ x_1^\prime = \frac{x_1}{r} ,\qquad x_2^\prime = \frac{x_2}{r} ,\qquad x_3^\prime = \frac{x_3}{r} ,\qquad \psi_1^\prime = \frac{\psi}{r} , $$
where $r^2=x_1^2+x_2^2+x_3^2$, maps solutions of the equation (\ref{Shre}) into solutions of the equation
$$ \Delta \psi^\prime + \frac{1}{(r\prime)^4}U(x_1^\prime,x_2^\prime,x_3^\prime)\psi^\prime =0 , $$
where $(r\prime)^2=  (x_1^\prime)^2+ (x_2^\prime)^2+  (x_3^\prime)^2$.
Similar arguments can easily be transferred to the $n$-dimensional Schrödinger equation.

\section{Parametric symmetries}
The search for contact germs preserving the differential ideals of equations is in general a difficult problem.
As is well known, Sophus Lie proposed the use of transformations acting on solutions of differential equations and forming local groups \cite{Ovs,Olver}.
In the paper \cite{KaptsovSFU} it was proposed to abandon some of the limitations of Lie theory. It was assumed  that transformations are of the form
$$ \tilde{x}_i = x_i + \sum_{k=1}^{\infty} a^k f_{ik}  \qquad
\tilde{u}^j = u^j + \sum_{k=1}^{\infty} a^k g_{jk}   , $$
where $f_{ik} , g_{jk}$ are analytic functions of $x_l, u^s_\alpha$ 
( $1\leq l\leq n$, $1\leq s\leq m$, $0\leq |\alpha|\leq p$).
Such transformations do not necessarily form a local group since they may be irreversible. 
In particular, it was shown in  \cite{KaptsovSFU} that Burgers' equation 
$$ u_t = u_{xx} +u u_x . $$
is invariant under the transformation
$$ t^{\prime} = t, \qquad x^\prime = x,\qquad u^\prime = u + 2D_x (\ln(1+a u))  , \qquad  a\in\mathbb{R} $$ 
and an infinite number of other transformations depending on higher derivatives 

 
 
 We now recall the construction from  \cite{KaptsovSFU} and
 assume for simplicity that $m=1$, $n=2$.
Let us consider a formal parametric mapping defined by power series in the parameter $a$
\begin{equation} \label{form}
	\tilde{x} = x +  \sum_{k=1}^{\infty} a^k x_k, \qquad 
	\tilde{y} = y +  \sum_{k=1}^{\infty} a^k y_k, \qquad \tilde{u} = u +  \sum_{k=1}^{\infty} 
	a^k u_k, 
\end{equation}
where $x_k, y_k, u_k$ are germs depending on $x, y, u, u_{10}, u_{01}$.
We further use the classical notation $p=u_{x},\ q=u_{y},\ r=u_{xx},\ s=u_{xy},\ t=u_{yy}$.

Suppose that the first prolongation of the formal mapping (\ref{form}) has the form
\begin{equation} \label{pq}
	\tilde{p}= p + \sum_{k=1}^{\infty} a^k p_k  , \qquad
	\tilde{q}= q + \sum_{k=1}^{\infty} a^kq_k , 
\end{equation}
where $p_k, q_k$ are germs depending on $x, y, u, p, q$.
We require that the Pfaffian equation
\begin{equation} \label{tilde u}
	d\tilde{u}-\tilde{p}d\tilde{x}-\tilde{q}d\tilde{y} =0
\end{equation}
was a consequence of the system
\begin{equation} \label{ du-}
	du = pdx + qdy, \quad dp = rdx +sdy, \quad dq = sdx +tdy.
\end{equation}
Substituting the series (\ref{form}), (\ref{pq}) into the equation (\ref{tilde u}) and using the system
(\ref{ du-}), we obtain the Pfaffian equation of the form
$$ Adx +Bdy =0 , $$
where $A, B$ are power series in $a$, all coefficients of which must be equal to zero.
Thus, collecting the terms containing $a^1$ into $A$ and $B$, we obtain
\begin{equation} \label{p1 q1}
	p_1 = D_x(u_1) - pD_x(x_1) - qD_x(y_1) , \quad
	q_1 = D_y(u_1) - pD_y(x_1) - qD_y(y_1) .
\end{equation}
These are the classical formulas for coefficients of the prolonged infinitesimal operator in Lie theory \cite{Ovs,Ibragimov}.

Equating coefficients of $a^2$ into $A$ and $B$ to zero yields
\begin{gather*} 
	p_2 =D_x(u_2) - pD_x(x_2)- p_1D_x(x_1)-qD_x(y_2) -q_1D_x(y_1) ,\\
	q_2 = D_y(u_2) - pD_y(x_2)- p_1D_y(x_1)-qD_y(y_2) -q_1D_y(y_1) . 
\end{gather*} 
Note that, in contrast to the group analysis of differential equations,  the germs $x_2, y_2, u_2$ are not expressed in terms of $x_1, y_1, u_1$.

The identical vanishing of coefficients of $a^n$ implies 
\begin{gather*} 
	p_n =D_x(u_n) - \sum_{i=0}^{n-1}(p_iD_x(x_{n-i})+q_iD_x(y_{n-i}) ) ,\\
	q_n = D_y(u_n) - \sum_{i=0}^{n-1}(p_iD_y(x_{n-i})+q_iD_y(y_{n-i}) ) , 
\end{gather*} 
with $p_0 = p,\ q_0=q$.

 Consider now the Pfaffian equations
\begin{equation} \label{TILDE p q}
d\tilde{p} = \tilde{r}d\tilde{x} + \tilde{s}d\tilde{y} , \qquad
d\tilde{q} = \tilde{s}d\tilde{x} + \tilde{t}d\tilde{y} ,
\end{equation}
where $\tilde{r}, \tilde{s}, \tilde{t}$ are represented by
 \begin{equation} \label{TILDE r s t}
\tilde{r}= r + \sum_{k=1}^{\infty} a^k r_k, \qquad
\tilde{s}= s + \sum_{k=1}^{\infty} a^k s_k, \qquad
\tilde{t}= t + \sum_{k=1}^{\infty} a^k t_k .
\end{equation}
The equations (\ref{TILDE p q}) must be consequences of the system
\begin{gather*}   
	dp = rdx + sdy   , \qquad    dq =sdx + tdy ,  \\
	dr = u_{30}dx +u_{21}dy, \quad  ds = u_{21}dx + u_{12}dy ,\quad
	dt = u_{12}dx +u_{03}dy .
\end{gather*} 
Substituting (\ref{pq}), (\ref{TILDE r s t}) into (\ref{TILDE p q}) and using the latter system, we get two expressions of the form
$$  A_1dx + B_1dy = 0, \qquad   A_2dx + B_2dy = 0          $$
where $A_1, B_1, A_2, B_2$ are power series in $a$, all coefficients of which must 
vanish identically.
Calculating the coefficients of these series in $a, a^2$, we have
\begin{gather*}
	r_1 = D_x(p_1)-rD_x(x_1)-sD_x(y_1) , \\ 
	s_1=D_y(p_1)-rD_y(x_1)-sD_y(y_1) ,\\
	t_1=D_y(q_1)-sD_y(x_1)-tD_y(y_1) , \\
	r_2 = D_x(p_2) -rD_x(x_2)-r_1D_x(x_1) - sD_x(y_2) -s_1D_x(y_1) ,\\ 
	s_2 = D_y(p_2) -rD_y(x_2)-r_1D_y(x_1) - sD_y(y_2) -s_1D_y(y_1) , \\ 
	t_2=  D_y(q_2) - sD_y(x_2) -s_1D_y(x_1) -tD_y(y_2) - t_1D_y(y_1) . 
\end{gather*} 
If necessary, the calculations are easy to continue.

As a first example, consider the equation
\begin{equation} \label{u_y}
  u_y - u^2 u_{xx} =0 .
\end{equation}
Let's find new symmetries of this equation. For this purpose one requires that the expression
\begin{equation} \label{TILDA u_y=}
	\tilde{u}_{\tilde{y}} - \tilde{u}^2 \tilde{u}_{\tilde{x}\tilde{x}} =
	\tilde{q} - \tilde{u}^2\tilde{r}
\end{equation}
lies in the differential ideal $<u_y - u^2 u_{xx} >$.
We first substitute expressions for $\tilde{q},\ \tilde{u},\ \tilde{r}$ into (\ref{TILDA u_y=}) and obtain a power series in $a$.
Equating coefficient of $a$ to zero leads to equation
\begin{equation} \label{q1-}
	q_1 - u^2 r_1 -2uru_1 =0  .
	\end{equation}
The left side of (\ref{q1-}) must lie in the differential ideal $<u_y - u^2 u_{xx} >$. Note that (\ref{q1-}) is the defining equation for the infinitesimal symmetries of the equation (\ref{u_y}). The equation (\ref{q1-}) reduces to a linear overdetermined partial differential system for the functions $x_1, y_1, u_1$.
If we assume that these functions depend on $x, y, u, u_x$, then the solution of this system has the form
\begin{equation} \label{x1=f}
	x_1 = f, \quad y_1 = c_1 y +c_2, \quad u_1 =u_x f +(c_3 x +c_4)u_x - u(c_3+ c_1/2) ,
\end{equation}
where $f$ is an arbitrary smooth function of $x, y, u, u_x$, and $c_1,\dots, c_4$ are arbitrary constants.

When we find the coefficient of $a^2$  and equate it to zero, we obtain the second defining equation for the functions $x_2, y_2, u_2$
\begin{equation} \label{q2--}
	q_2 - u^2 r_2 -2ur_1u_1 - r u_1^2 - 2ru u_2=0 ,
\end{equation}
which should also be a differential consequence of the equation (\ref{u_y}). 
If the function $f$ in (\ref{x1=f}) vanishes identically, then it is not possible to find non-trivial solutions to the second defining equation (\ref{q2--}).

Suppose that $f\neq 0$ and $c_1=\cdots c_4 =0$, then a nontrivial solution of  the equation (\ref{q2--}) can exist if 
$$ f = u\sqrt{b_1 y + b_2 -2y_2} ,\qquad  
 u_2= (x_2 +b_3 x + b_4) u_x - (b_3 +b_4/4)u , $$
where $x_2, y_2$ are arbitrary functions of $x, y, u, u_x$ and $b_1, b_2, b_3, b_4$ are arbitrary constants.
Assuming $x_2=y_2=b_1=b_3=b_4=0$, and $b_2=1$, we obtain the mapping
\begin{equation} \label{sym1}
	\tilde{x} = x +au, \qquad \tilde{y} = y, \qquad \tilde{u} = u + auu_x ,
\end{equation}
generating the non Lie symmetry of the equation (\ref{u_y}).
Repeated application of the map (\ref{sym1}) leads to a symmetry depending on $u_{xx}$. This process can be repeated any number of times
 and  new symmetries can be obtained.

Let us now consider a system of linear partial differential equations
\begin{equation} \label{Lu=0}
	L(x, \partial)u = 0, \qquad L(x,\partial) = \sum_{|\alpha|\geq 0}^{k} A_\alpha \partial^\alpha ,
\end{equation}
where $u=(u^1,\dots,u^m)$, $\partial^\alpha = \frac{\partial^\alpha }{\partial x_1^{\alpha_1}\cdots\partial x_n^{\alpha_n}} $, $\alpha= (\alpha_1,\dots,\alpha_n) $, and $A_\alpha$ are $m\times m$ are matrices depending on $x=(x_1,\dots,x_n)$, .
Suppose this system admits a symmetry operator written in the canonical form \cite{Ibragimov, Olver}
$$ X = \sum_{i=1}^{m}\eta_i \frac{\partial }{\partial u^i} +
 \sum_{\substack{1\leq j\leq m \\ \alpha\in \mathbb{N}^n }} D^\alpha\eta_i 
 \frac{\partial }{\partial u^i_\alpha}  $$
 Then the collection $\eta =(\eta_1,\dots,\eta_m)$ must satisfy the condition
$$ L(x,D)\eta \in<L(x,\partial)u> .$$
Consequently, if $u$ is a solution to the system (\ref{Lu=0}), then $\eta$ is also a solution to this system.

Consider now the one-dimensional gas dynamics equations 
\begin{gather*}
	u_t +uu_x +\rho^n\rho_x =0    ,\qquad \rho_t +u\rho_x +u_x\rho =0 ,
\end{gather*}
where $u$ is the gas velocity, and $\rho$ is the density. We assume that 
entropy remains constant and $n\in\mathbb{R}$. 
We use the hodograph transformation \cite{Whitham}, and interchange the roles of dependent and independent variables. This leads to a system 
\begin{equation} \label{Lsys}
 x_\rho - u t_\rho + \rho^n t_u =0 ,  \qquad    
 x_u -u t_u +\rho t_\rho =0  . 
\end{equation}

One of the admited infinitesimal operators of the last system in canonical form is 
$$  (bt+at_u)\frac{\partial }{\partial t} + (bx + a(x_u -t))\frac{\partial }{\partial x}  ,\qquad \forall a,b\in\mathbb{R}   . $$
So the transformation
$$ \tilde{u}= u ,\quad \tilde{\rho}= \rho ,\quad \tilde{t}=bt+at_u,\quad \tilde{x}= bx + a(x_u -t) $$
maps solutions of a system (\ref{Lsys}) into solutions of the same system, i.e. this transformation is a symmetry.
This map in the original variables has the form
$$ \tilde{u}= u ,\quad \tilde{\rho}= \rho ,\quad \tilde{t}=bt + \frac{a\rho_x}{\Delta }\ ,
\quad \tilde{x}= bx - a\frac{\rho_t +t}{\Delta }\ ,$$
with $\Delta = u_t\rho_x - u_x\rho_t$.
It is a symmetry of the gas dynamics equations. In this way we can construct a series of symmetries.

\section{Conclusion}

The paper proposes a new approach to the study of symmetries of differential equations.
To present the classical theory and its generalizations, it turns out to be convenient to use germs of functions and differential forms.
Parametric symmetries are of great interest for further research, since this makes it possible to use parameter expansion and sequentially solve the defining equations.
Parametric families of germs are, in general, not invertible and can depend on derivatives of finite order.
 To show the effectiveness of this approach  
 it is necessary to find new examples of differential equations admitting parametric symmetries. 
 
This work is supported by the Krasnoyarsk Mathematical Center and financed by the Ministry of Science and Higher Education of the Russian Federation in the framework of the establishment and development of regional Centers for Mathematics Research and Education (Agreement No. 075-02-2024-1378).

\end{document}